\documentclass[twocolumn]{aastex62}

\usepackage{graphicx}
\usepackage{amsmath}
\usepackage{float}
\usepackage{natbib}
\usepackage{longtable}
\usepackage{color}

\graphicspath{{./}{figures/}}

\shorttitle{NS--NS/BH Mergers in GCs}
\shortauthors{Ye et al.}

\begin{document}

\title{On the Rate of Neutron Star Binary Mergers from Globular Clusters}

\author[0000-0001-9582-881X]{Claire S.\ Ye}
\affil{ Department of Physics \& Astronomy, Northwestern University, Evanston, IL 60208, USA}
\affil{ Center for Interdisciplinary Exploration \& Research in Astrophysics (CIERA), Northwestern University, Evanston, IL 60208, USA}
\correspondingauthor{Claire S. Ye}
\email{shiye2015@u.northwestern.edu}

\author[0000-0002-7374-935X]{Wen-fai Fong}
\affil{ Department of Physics \& Astronomy, Northwestern University, Evanston, IL 60208, USA}
\affil{ Center for Interdisciplinary Exploration \& Research in Astrophysics (CIERA), Northwestern University, Evanston, IL 60208, USA}

\author[0000-0002-4086-3180]{Kyle Kremer}
\affil{ Department of Physics \& Astronomy, Northwestern University, Evanston, IL 60208, USA}
\affil{ Center for Interdisciplinary Exploration \& Research in Astrophysics (CIERA), Northwestern University, Evanston, IL 60208, USA}

\author[0000-0003-4175-8881]{Carl L.\ Rodriguez}
\affil{Harvard Institute for Theory and Computation, 60 Garden St, Cambridge, MA 02138, USA}

\author[0000-0002-3680-2684]{Sourav Chatterjee}
\affil{Tata Institute of Fundamental Research, Homi Bhabha Road, Mumbai 400005, India}
\affil{ Center for Interdisciplinary Exploration \& Research in Astrophysics (CIERA), Northwestern University, Evanston, IL 60208, USA}

\author[0000-0002-7330-027X]{Giacomo Fragione}
\affil{ Department of Physics \& Astronomy, Northwestern University, Evanston, IL 60208, USA}
\affil{ Center for Interdisciplinary Exploration \& Research in Astrophysics (CIERA), Northwestern University, Evanston, IL 60208, USA}

\author[0000-0002-7132-418X]{Frederic A.\ Rasio}
\affil{ Department of Physics \& Astronomy, Northwestern University, Evanston, IL 60208, USA}
\affil{ Center for Interdisciplinary Exploration \& Research in Astrophysics (CIERA), Northwestern University, Evanston, IL 60208, USA}

\begin{abstract}
The first detection of gravitational waves from a neutron star -- neutron star (NS--NS) merger, GW170817, and the increasing number of observations of short gamma-ray bursts (SGRBs) have greatly motivated studies of the origins of NS--NS and neutron star -- black hole (NS--BH) binaries. We calculate the merger rates of NS--NS and NS--BH binaries from globular clusters (GCs) using realistic GC simulations with the \texttt{CMC} cluster catalog. We use a large sample of models with a range of initial numbers of stars, metallicities, virial radii and galactocentric distances, representative of the present-day Milky Way GCs, to quantify the inspiral times and volumetric merger rates as a function of redshift, both inside and ejected from clusters. We find that over the complete lifetime of most GCs, stellar BHs dominate the cluster cores and prevent the mass segregation of NSs, thereby reducing the dynamical interaction rates of NSs so that at most a few NS binary mergers are ever produced. We estimate the merger rate in the local universe to be $\sim\rm{0.02\,Gpc^{-3}\,yr^{-1}}$ for both NS--NS and NS--BH binaries, or a total of $\sim 0.04$~Gpc$^{-3}$~yr$^{-1}$ for both populations. These rates are about 5 orders of magnitude below the current empirical merger rate from LIGO/Virgo. We conclude that dynamical interactions in GCs do not play a significant role in enhancing the NS--NS and NS--BH merger rates.
\end{abstract}


\section{INTRODUCTION}\label{sec:intro}
Since the discovery of the first neutron star -- neutron star (NS--NS) binary, PSR B1913+16 \citep{hulse1975discovery}, 20 NS--NSs have been observed in the radio band in our Milky Way Galaxy alone (\citealp[][and references therein]{tauris2017formation}; \citealp{martinez2017pulsar,cameron2018high,lynch2018green,stovall2018palfa,Ridolfi2019upgraded}). More recently, the first gravitational-wave signal from a NS--NS merger, GW170817, was detected by the Advanced LIGO/Virgo network \citep{abbott2017gw170817}. GW170817 was followed by the detection of a short gamma-ray burst (SGRB), one of a class of explosions long suspected to originate from NS--NS and/or neutron star -- black hole (NS--BH) mergers \citep[e.g.,][]{narayan1992grb,berger2014short}. Two primary formation channels have been suggested for NS--NS and NS--BH mergers: isolated binary evolution of massive stars and dynamical formation in dense stellar environments such as globular clusters (GCs).

Previous studies have shown that merging black hole -- black hole (BH--BH) binaries are formed at substantial rates in GCs, high enough to explain the LIGO/Virgo detection rate \citep{rodriguez2015binary,rodriguez2016binary,fragione2018black,Hong2018binary,rodriguez2018posta,rodriguez2018postb,samsing2018black,Choksi2019star,kremer2019post,samsing2019gravitational}. The reason is that dynamical interactions in GCs greatly boost the formation and merger rates of BH--BH binaries \citep[e.g.,][]{rodriguez2016binary}. This naturally leads to the question of whether dynamics in GCs could similarly contribute to the NS--NS and NS--BH merger rates. On the one hand, there are many more binaries in the field than in GCs that can become NS--NS or NS--BH binaries. On the other hand, dynamical interactions in GCs are very efficient at forming compact object binaries with NSs, such as low-mass X-ray binaries (LMXBs) and millisecond pulsar (MSP) binaries \citep[e.g.,][]{clark1975x,Pooley2003dynamical,Bahramian2013stellar,ye2019millisecond}.

There are ongoing debates about the contribution of NS--NS/NS--BH mergers from GCs to the overall merger rates in the Universe. \cite{grindlay2006short} and \cite{lee2010short} estimated that the merger rate from NS--NS binaries formed dynamically in GCs with properties similar to M15 (massive and core-collapsed) can account for $10-30\%$ or more of SGRBs. \cite{guetta2009short} calculated a very high NS--NS merger rate from GCs by fitting the SGRB luminosity function and observed redshift distribution. \cite{andrews2019double} showed that the binary properties of a few NS--NS binary pulsars in the Galactic field are difficult to explain with isolated binary evolution, and instead suggested that some NS--NS binaries must be formed in GCs through stellar dynamics. Observationally, studies of SGRBs have found large offsets of these sources relative to the centers of their host galaxies, suggesting that their progenitors could have been in GCs and subsequently ejected to the outer halos of galaxies \citep[e.g.,][]{fong2013locations,berger2014short}.

On the other hand, several studies have suggested that the NS--NS merger rate from GCs is low compared to the field. \cite{bae2014compact} used direct $N$-body simulations to estimate a merger rate of NS--NS binaries ejected from GCs of less than $0.1\%$ of the overall NS--NS merger rate. This is in agreement with early inferred rates from the first three binary pulsars observed in the Milky Way \citep{phinney1991rate}. \cite{belczynski2018origin} computed a set of GC models assuming small NS natal kicks and using the \texttt{MOCCA} code \citep[e.g.,][]{giersz2013mocca}, and derived a NS--NS merger rate from GCs about 4 orders of magnitude lower than the merger rate from isolated binary evolution. These results are consistent with the latest deep {\it Hubble Space Telescope\/} imaging of the location of GW170817, which has definitively ruled out a GC as a merger site for this event \citep{fong2019optical}.

We are aware of only one past study that attempted to estimate the NS--BH merger rate from GCs: \cite{clausen2013black} followed the evolution of NS--BH binaries undergoing binary--single stellar interactions in static background cluster models. Compared to the current LIGO/Virgo merger rate upper limit (610~$\rm{Gpc^{-3}\,yr^{-1}}$) for NS--BHs, their estimated merger rate from GCs (0.01--0.17~$\rm{Gpc^{-3}\,yr^{-1}}$) appears negligible.

Here we compute the NS--NS and NS--BH merger rates from GCs using Monte Carlo simulations of cluster dynamics. For the first time, we use a large sample of realistic models representing Milky Way GCs with different initial numbers of stars, metallicities, virial radii and galactocentric distances. In Section~\ref{sec:models}, we describe the methods we use to model GCs. In Section~\ref{sec:results}, we discuss how stellar BHs control the dynamics of NSs, and we quantify the NS--NS and NS--BH binary formation times, inspiral times, and volumetric merger rates as a function of redshift. In Section~\ref{sec:previousstudy}, we compare our results with those of previous studies. In Section~\ref{sec:discussions}, we summarize our findings and discuss some caveats.

\section{MODELING GLOBULAR CLUSTERS}\label{sec:models}
This work is based on a set of 144 GC models computed with our \texttt{Cluster Monte Carlo} code (\texttt{CMC} cluster catalog), described in more detail in \citet{kremer2019grid}. \texttt{CMC} is a H\'{e}non-type Monte Carlo code \citep{henon1971monte,henon1971montecluster} that has been developed over many years \citep{Joshi_2000,Joshi_2001, Fregeau_2003,fregeau2007monte, Chatterjee_2010,Umbreit_2012, Pattabiraman_2013,Chatterjee_2013b, rodriguez2018posta}. It incorporates all the relevant physics for GC evolution, including two-body relaxation, three-body binary formation, strong three- and four-body interactions, and some post-Newtonian effects \citep{rodriguez2018posta}. Updated versions of \texttt{SSE} \citep{hurley2000comprehensive} and \texttt{BSE} \citep{hurley2002evolution} are used to model the evolution of single stars and binary stars, respectively. The \texttt{Fewbody} package is used to directly integrate all three- and four-body gravitational encounters \citep{fregeau2004stellar,fregeau2007monte}, with some post-Newtonian effects included \citep{antognini2014rapid,amaro2016relativistic}.

Table~\ref{tab:modelgrid} (Appendix) shows the properties of all of our models (henceforth we refer to these models as ``realistic'' as opposed to the extreme limiting case described below), which are allowed to evolve up to 14~Gyr. Their initial conditions span wide ranges, with  initial number of stars $N = 2\times10^5$, $4\times10^5$, $8\times10^5$, and $1.6\times10^6$, initial virial radius $r_v = 0.5$, 1, 2, and $4\,$pc, metallicity $Z = 0.01$, 0.1, and $1\,\rm{Z_{\odot}}$, and galactocentric distance $r_g = 2$, 8, and $20\,\rm{kpc}$ (used to set the tidal boundary of the cluster). All models have a $5\%$ initial binary fraction for all stars and a King concentration parameter $W_0 = 5$ \citep{Heggie_Hut_Book}. We assume that the natal kicks for NSs formed in both core-collapse supernovae (CCSNe) and electron-capture supernovae (ECSNe) are sampled from a Maxwellian distribution with velocity dispersion $\sigma_{\rm{CCSN}}=265\,\rm{km\,s^{-1}}$ \citep{hobbs2005statistical} and $\sigma_{\rm{ECSN}}=20\,\rm{km\,s^{-1}}$ \citep{Kiel_2008}, respectively. We assume that BHs are born with fallback kicks: their natal kicks are drawn from the same Maxwellian distribution as for the CCSNe NSs, but with the velocity dispersion reduced by the amount of the fallback material \citep[see][for more details]{Fryer2012explosion,morscher2015dynamical}. We choose these initial parameters for the models so that they evolve to represent fully the present-day GCs in the Milky Way \citep{kremer2019grid}.

We have shown in \cite{ye2019millisecond} that the production of NS binaries is more efficient when most of the BHs have first been ejected out of the cluster. Therefore, as a way of obtaining an extreme upper limit for the NS--NS and NS--BH merger rates from GCs, we added one ``extremely optimistic model,'' which retains very few BHs. In this model, we truncate the upper end of the initial mass function (IMF; \citealp{kroupa2001variation}) at 30~$\rm{M_{\odot}}$, and we take $\sigma_{\rm{BH}} = 2000\,\rm{km\,s^{-1}}$; instead we take $\sigma_{\rm{NS}} = 20\,\rm{km\,s^{-1}}$, so that many NSs are retained. Furthermore, to enhance the formation rates of compact object binaries, we also assume an initial $100\%$ binary fraction for massive stars $>15\rm{M_{\odot}}$ (while the binary fraction remains $5\%$ for lower-mass stars). For this model we set $N=8\times10^5$, $r_v=0.5\,$pc, $Z = 0.05\,\rm{Z_{\odot}}$, $r_g = 8\,\rm{kpc}$ and $W_0 = 5$, such that this model goes into deep core collapse at late times. We evolve this model for 12~Gyr.

\section{NEUTRON STAR BINARY MERGER PROPERTIES AND RATES}\label{sec:results}
\subsection{The Role of Black Holes in Neutron Star Dynamics}\label{subsec:nsdyn}
\begin{figure*}
\begin{center}
\includegraphics[width=\textwidth]{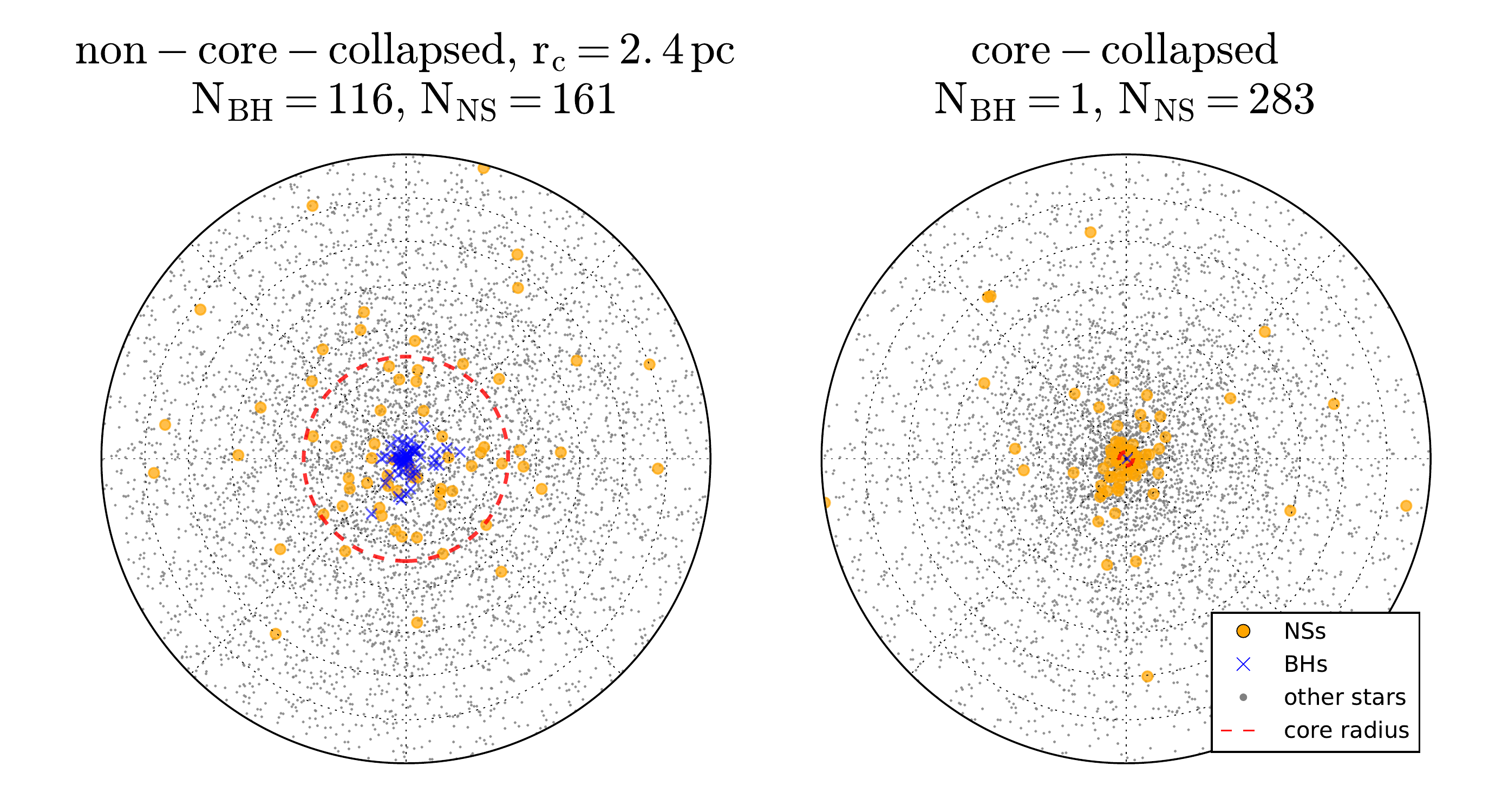}
\vspace{-0.3in}
\caption{Projected radii of stars out to 7 pc in two typical GC models. On the left is a non-core-collapsed cluster (model N8-RV2.0-RG8-Z0.1 in Table~\ref{tab:modelgrid}) with core radius 2.4~pc at 12~Gyr. On the right is a core-collapsed cluster (model N8-RV0.5-RG8-Z0.1 in Table~\ref{tab:modelgrid}). The blue crosses and orange dots show the BHs and NSs, respectively. The grey dots show all other types of stars. Core radii are shown by dashed red circles. \label{fig:2dproj}}
\end{center}
\end{figure*}

Figure~\ref{fig:2dproj} shows two typical GCs from our realistic models at 12~Gyr. There are some clear differences between these two models: the core-collapsed cluster has only few BHs remaining at late times (and just 1 retained at 12~Gyr). In contrast, the non-core-collapsed cluster retains many BHs (116 BHs at 12~Gyr) and has a large core radius of about 2~pc. The NSs in the core-collapsed cluster mass-segregate much further towards the cluster center than in the non-core-collapsed cluster, which can enhance the NS--NS and NS--BH binary formation and merger rates. Thus, it is not surprising that the first and only confirmed NS--NS binary in a GC is the binary pulsar PSR2127+11C in M15 \citep{anderson1990discovery,prince1991timing}, a prototypical core-collapsed cluster \citep{Harris2010catalog} \footnote{There is a candidate NS--NS binary pulsar in NGC~6544 \citep{lynch2012timing}, which is also a core-collapsed cluster \citep{Harris2010catalog}. Another possible NS--NS binary was also recently found in NGC 1851 (massive and with a very small core radius; \citealp{Ridolfi2019upgraded}).}.

These differences between core-collapsed and non-core-collapsed clusters are caused by the influence of the BHs on the evolution of their host clusters and on the dynamics of the NSs \citep[for recent studies, see][]{kremer2019initial,kremer2019role,ye2019millisecond}. BHs dominate the cluster cores due to mass segregation. Their dynamical interactions and the resulting heating of the cluster cores (``BH burning"; \citealp[and the references therein]{kremer2019role}) inhibit the NSs from further mass-segregating to the cluster cores (Fig.~\ref{fig:2dproj}), where stellar densities are highest and stars experience dynamical interactions. Only after most of the BHs are ejected out of the cluster (taking at least a few Gyr; \citealp{kremer2019initial}) can the NSs move to the center and interact to form systems such as MSPs, NS--NS binaries and NS--BH binaries \citep[e.g.,][]{fragione2018neutron,ye2019millisecond}. This was also pointed out in \citet{zevin2019can} which explored NS--NS mergers in GCs in the context of $r$-process enrichment, also using \texttt{CMC} models.

\subsection{Merger Statistics}\label{subsec:statistics}
We define two main formation mechanisms in our simulations: {\it primordial\/} NS--NS and NS--BH binaries, which evolve from primordial massive binaries in the cluster (but may have experienced weak dynamical interactions during their lifetime), and {\it dynamical\/} NS--NS and NS--BH binaries, which are assembled through dynamical exchange interactions. We do not take into account the ejected binaries (NS--main-sequence or NS--giant binaries) that might produce NS--NS or NS--BH mergers at later times. In all our models, we find $\sim 90$ of such binaries with a companion mass above $7\,M_{\odot}$. Almost all of these binaries are primordial and ejected from their cluster as a result of the first SN kick. However, only a very small fraction of these primordial binaries are expected to eventually produce mergers \citep[e.g.,][]{belczynski2018origin,Chruslinska2018double} and therefore we neglect them in our analysis, which focuses instead on dynamically produced binaries.

First, we consider all NS--NS binaries in the realistic models (Table~\ref{tab:modelgrid}). In total, there are 64 NS--NS mergers in 119 realistic models that survived to $12\,$Gyr (25 of the models evolved to complete disruption). Most of the NS--NS binaries are primordial (about $70\%$). We find that $83\%$ of all NS--NS mergers are in binaries ejected from their cluster (i.e., merging in the field), and only $17\%$ merge inside clusters. Of the ejected merging binaries, $83\%$ are primordial (ejected immediately at formation due to large natal kicks), and $17\%$ are dynamical. In contrast, all binaries that merge in clusters are dynamically assembled.

For NS--BHs, we find 31 mergers in 119 realistic models, about half the number of NS--NS mergers. In contrast to NS--NS mergers, most of the merging NS--BH binaries form in dynamical encounters (about $80\%$). About $35\%$ merge outside clusters, and $65\%$ merge inside clusters. Among ejected NS--BH binaries that merge, $55\%$ are primordial. Almost all of the in-cluster mergers are dynamical. Compared to NS--NS mergers, a larger fraction of NS--BH binaries merge inside clusters. This is expected, since NS--BH binaries are more massive than NS--NS binaries and therefore harder to eject, and because they tend to form late in the evolution of clusters when few other BHs remain. There is also a larger fraction of dynamical NS--BH binaries than dynamical NS--NS binaries (80\% vs. 30\%). As expected (Sec.~\ref{subsec:statistics}), almost all of the dynamical NS--NS and NS--BH binaries are from core-collapsed clusters.

We also note that many of the merging binaries contain active pulsars at the time of merger. In total, 43 of 64 NS--NSs merging in our models ($67\%$) contain either a MSP or a young pulsar (with higher magnetic field and longer spin period than a typical MSP; see \citealp{ye2019millisecond}). Meanwhile 8 of 31 NS--BH mergers ($26\%$) contain at least one active pulsar.

In the extremely optimistic model, there are 139 NS--NS mergers and 39 NS--BH mergers. In contrast to the realistic models, about $94\%$ of the NS--NS binaries are dynamical, and only about $40\%$ of the NS--NS binaries are ejected and merge outside of the cluster. While similar to the realistic models, we find that all of the NS--BH binaries are dynamical. About $26\%$ of them are ejected and merge outside the cluster. Furthermore, all of the primordial NS--NS binaries merge outside the cluster.

\begin{figure}
\begin{center}
\includegraphics[width=\columnwidth]{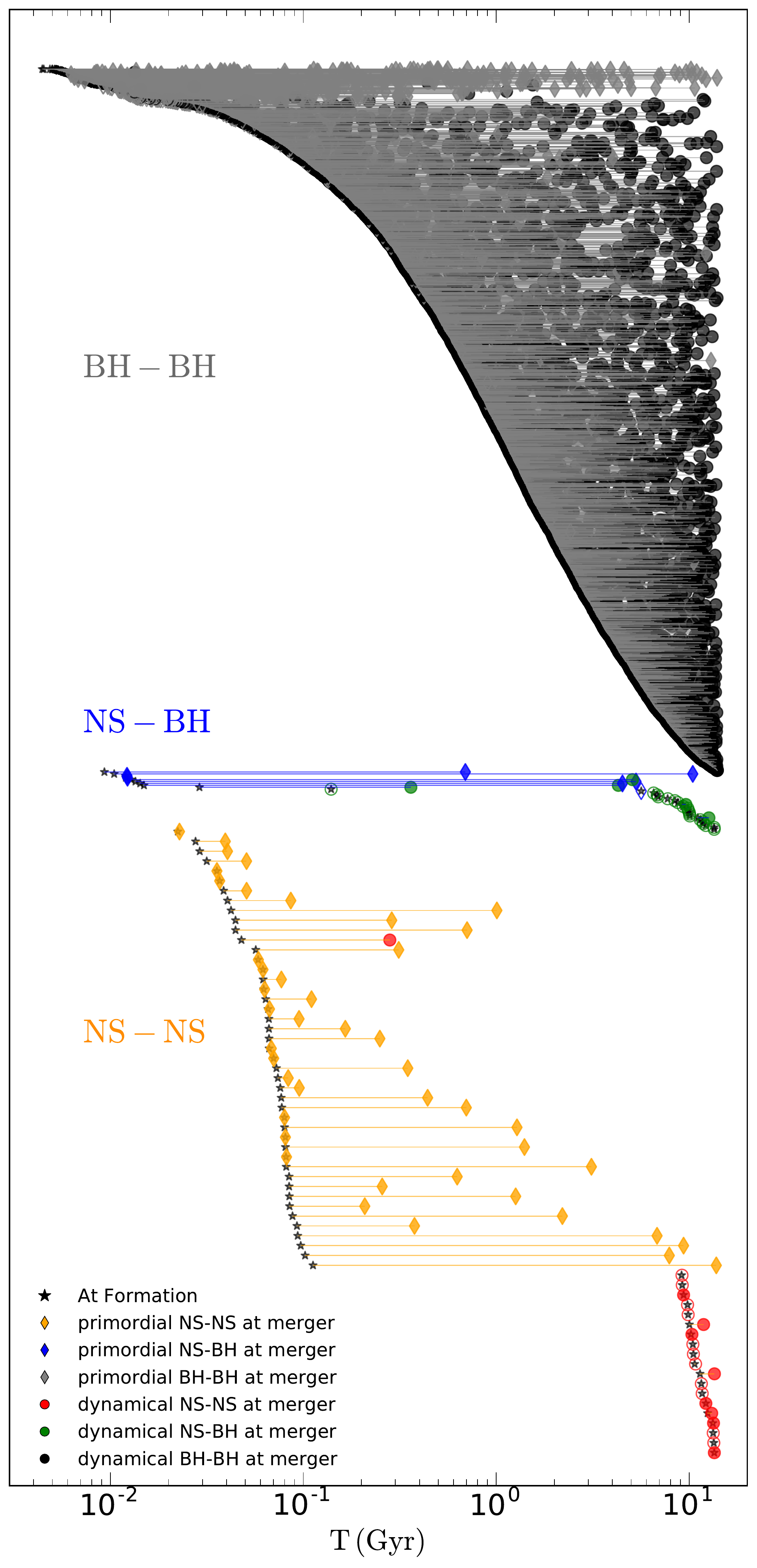}
\vspace{-0.25in}
\caption{Formation and merger times of NS--NS, NS--BH and BH--BH binaries (relative to the time of birth of the cluster) in all of our realistic models surviving to the present. Only systems that merge within a Hubble time are shown, ordered by formation time within each group along the y-axis. Lines connect the same systems. Black stars mark the formation time (or time of the last interaction). Diamonds and circles mark the merger time. Different colors and shapes show different formation channels. Orange, blue and grey diamonds show primordial NS--NS, NS--BH and BH--BH binaries, respectively. Red, green and black circles show dynamical NS--NS, NS--BH and BH--BH binaries, respectively. Open symbols indicate mergers in clusters and filled symbols indicate mergers in the field. It is immediately clear that GCs can produce a large number of BH--BH mergers (10824 in 119 models), but comparatively few NS--NS (64) and NS--BH (31) mergers, and that dynamics play a key role in determining when NS--NS and NS--BH mergers occur. \label{fig:alltimes}}
\end{center}
\end{figure}

\subsection{Formation, Merger and Inspiral Times}\label{subsec:merger_time}
\begin{figure}
\begin{center} 
\includegraphics[width=\columnwidth]{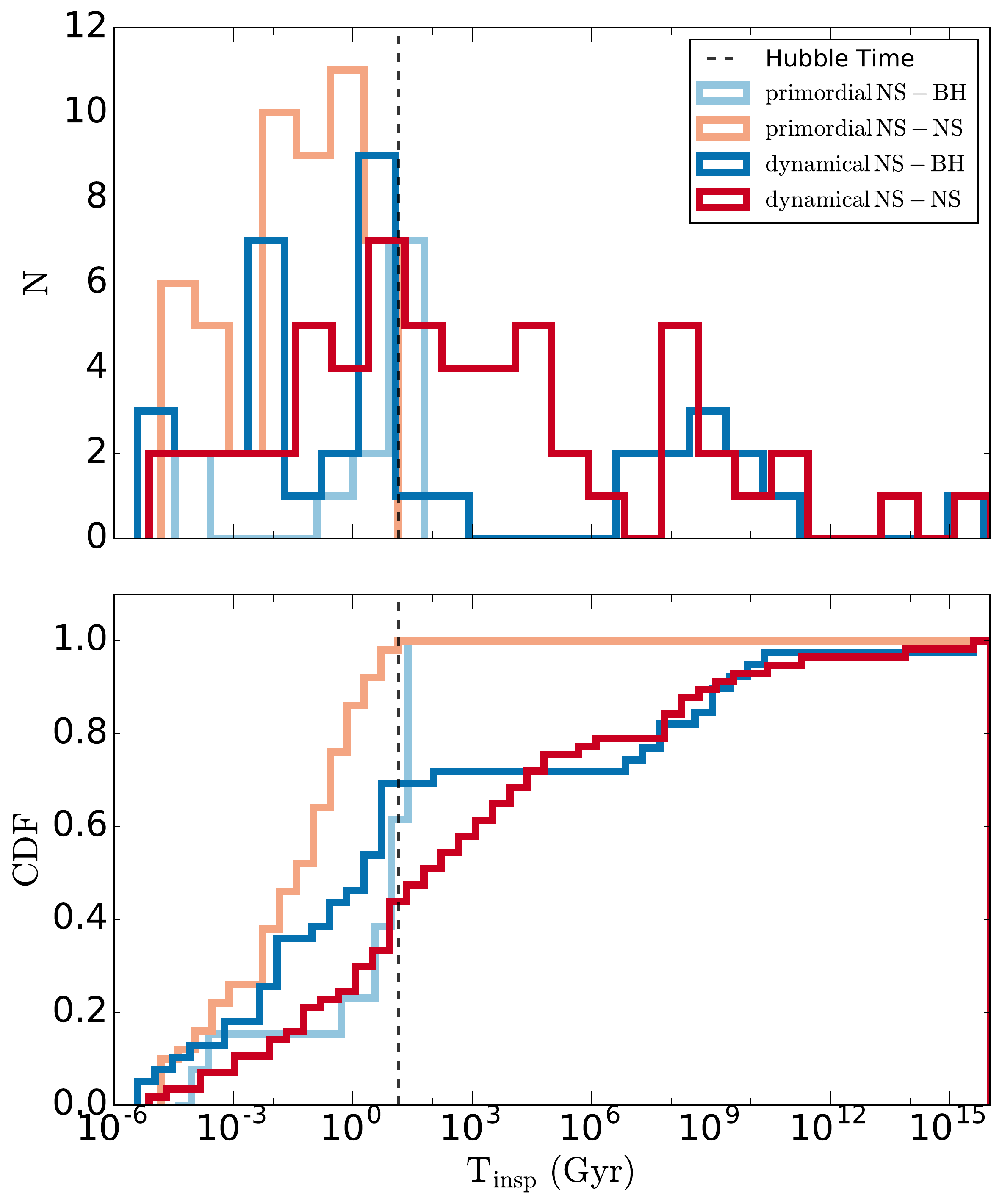}
\vspace{-0.2in}
\caption{Inspiral time distributions of all NS--NS and NS--BH binaries in realistic models. The Hubble time is shown by the dashed black line. The majority of the primordial NS--NS and NS--BH binaries (light red and light blue lines, respectively) have inspiral times less than a Hubble time, while the inspiral times of the dynamical NS--NS and NS--BH binaries (dark blue and dark red lines) span a wider range. \label{fig:tinspiral}}
\end{center}
\end{figure}

We extract the time of formation and merger (relative to the birth of the GC) from our simulations of BH--BH, NS--BH and NS--NS binaries in our realistic models (Fig.~\ref{fig:alltimes}), excluding those merging beyond a Hubble time. For the primordial NS--NSs (orange diamonds), we see that most of them merge early (before 3~Gyr) and outside of the cluster. This can be naturally explained since most of them are ejected at formation due to supernova kicks and their inspiral times peak at about 100~Myr (Fig.~\ref{fig:tinspiral}). In contrast, more than half of the dynamical NS--NS mergers occur at around 10~Gyr and inside GCs. This is because the timescale for the NSs to take part in dynamical interactions and form NS--NS binaries is at least several Gyr, after most of the BHs have been ejected from the cluster \citep{kremer2019initial}. Frequent dynamical encounters of newly-formed NS--NS binaries in the GC cores can then harden them and quickly lead them to merge. As a result, most dynamical NS--NS binaries merge at low redshift ($z\lesssim0.5$) and most of the primordial NS--NS binaries merge at high redshift ($z\gtrsim2$).

Similarly, most of the dynamical NS--BH mergers form and merge at around 10~Gyr, for the same reason as discussed above for NS--NS mergers. The primordial NS--BH binaries also form early in GCs, but some of them merge at late times because of the large orbital period ($>5$~days) they acquire when ejected (Fig.~\ref{fig:tinspiral}).

Figure~\ref{fig:tinspiral} shows the inspiral times (from formation to merger) for all NS--NS and NS--BH binaries, including those merging beyond a Hubble time. We simply integrate the equations in \cite{peters1964gravitational} to calculate the inspiral times of ejected systems. The inspiral times of the primordial NS--NS and NS--BH binaries are mainly determined by binary evolution and their binary properties at ejection. Most of them merge within a Hubble time. This is because in order to stay bound in spite of the large supernova kicks, these binaries must undergo common-envelope evolution, leading to very tight orbits. The dynamical NS--NSs and NS--BHs, however, have a wider inspiral time distribution since dynamical interactions can produce a wider range of orbital periods and eccentricities.

\newpage
\subsection{Merger Rates}\label{subsec:rates}
\begin{deluxetable*}{ccccc}[t]
\tabletypesize{\footnotesize}
\tablewidth{0pt}
\tablecaption{Derived Merger Rates from Globular Clusters at $z<0.1$\label{tab:modelrates}}
\tablehead{
\colhead{Models} & \colhead{Systems} & \colhead{$\rho_1 (0.33\, \rm{Mpc^{-3}})$} & \colhead{$\rho_2 (0.77\, \rm{Mpc^{-3}})$} &\colhead{$\rho_3 (2.31\, \rm{Mpc^{-3}})$}\\
\colhead{} & \colhead{} & \colhead{$\rm{Gpc^{-3}yr^{-1}}$} & \colhead{$\rm{Gpc^{-3}yr^{-1}}$} & \colhead{$\rm{Gpc^{-3}yr^{-1}}$}
}
\startdata
Realistic & NS--NS (NS--BH) & $0.009$ ($0.009$) & $0.022$ ($0.020$) & $0.065$ ($0.060$)\\
Realistic & DYN NS--NS (NS--BH) & $0.008$ ($0.008$) & $0.019$ ($0.018$) & $0.057$ ($0.055$)\\
Realistic & PRIM NS--NS (NS--BH) & $0.001$ ($0.001$) & $0.003$ ($0.002$) & $0.008$ ($0.005$)\\
Extremely Optimistic & NS--NS (NS--BH) & $3.6$ ($0.8$) & $8.5$ ($1.8$) & $25.5$ ($5.5$)\\
\enddata
\tablecomments{Estimated merger rates for the extremely optimistic model and the realistic models from the main grid. $\rho_1$, $\rho_2$ and $\rho_3$ are different assumed GC densities in the local universe. NS--NS or NS--BH denote the total merger rates for NS--NS or NS--BH binaries, including both primordial and dynamical binaries. Prefix ``DYN" denotes dynamically assembled binaries, and prefix ``PRIM" denotes primordial binaries.}
\end{deluxetable*}

We can estimate the merger rates of NS--NS and NS--BH binaries from our models as a function of redshift adopting a similar approach to that used in \cite{oleary2006binary} and \cite{rodriguez2016binary}. The comoving merger rate is calculated as 
\begin{equation} \mathcal{R}(z)=\frac{dN(z)}{dt}\times\rho_{GC},\end{equation} 
where $dN(z)/dt$ is the number of mergers per unit time per GC at a given redshift, and $\rho_{GC}$ is the volumetric number density of GCs in the local universe. All mergers in our models are assigned to one of 400 time bins uniformly covering the 14~Gyr of dynamical evolution. We computed $dN(z)/dt$ by summing the total number of mergers in each bin, and averaging the numbers over the bin width and the total number of models in the calculation. We use three different GC densities in the local universe for the rate calculation: $\rho_{\rm{GC}}$=0.33~$\rm{Mpc^{-3}}$, 0.77~$\rm{Mpc^{-3}}$ and 2.31~$\rm{Mpc^{-3}}$ \citep[][and references therein]{rodriguez2016binary}.

The cumulative merger rate is calculated as
\begin{equation} R_c(z)=\int_{0}^{z}\mathcal{R}(z') \times\frac{dV_c}{dz'}\times(1+z')^{-1}dz'.\end{equation} 
Here $\mathcal{R}(z')$ is the comoving merger rate from equation~(1), $dV_c/dz'$ is the comoving volume at redshift $z'$ and $(1+z')^{-1}$ is a correction to account for time dilation.

The differences in ages for GCs with different metallicities are also taken into account. We use the age distributions in \cite{el2018formation}, and the ages are divided into three bins with metallicity ranges $Z\leq0.00065$, $0.00065< Z \leq 0.0065$ and $Z>0.0065$, respectively. The age distributions peak at around 13, 11 and 9~Gyr for the three bins. The three metallicites of our realistic models are roughly at the center of each bin. Applying the age distributions to the models (which are all allowed to evolve for 14~Gyr) gives them more realistic ages, and the times of mergers are adjusted accordingly. To convert time in Gyr to redshift, we assume $H_0 = 69.6\,\rm{km\,s^{-1}\,Mpc^{-1}}$, $\Omega_M = 0.286$, and $\Omega_{\Lambda} = 0.714$ \citep{bennett20141}.

\begin{figure}
\begin{center}
\includegraphics[width=\columnwidth]{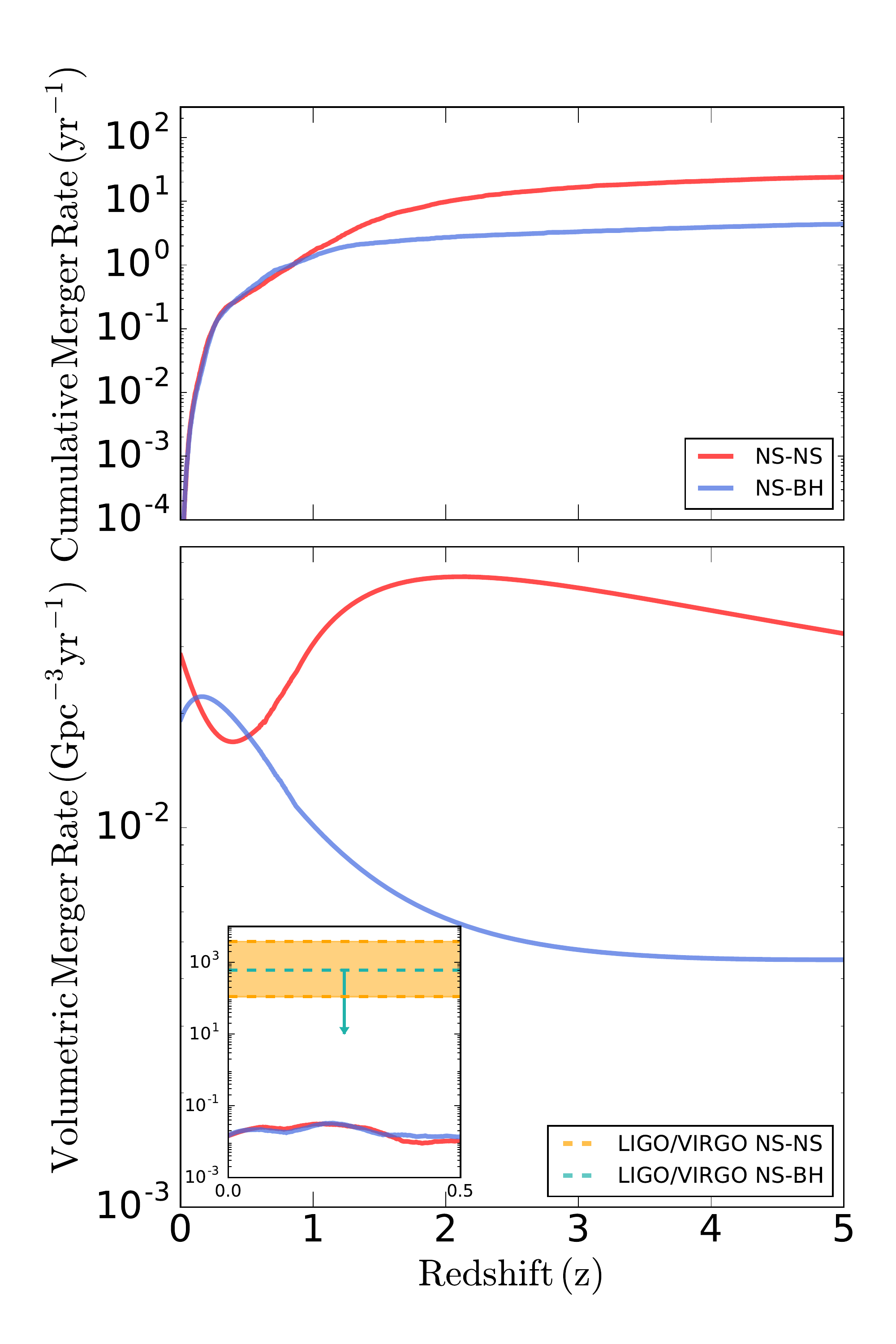}
\vspace{-0.3in}
\caption{Merger rates as a function of redshift for the realistic models. The upper panel shows the cumulative merger rates per year, and the lower panel shows the merger rate densities per $\rm{Gpc^{3}}$ per year. Red and blue curves are for our model NS--NS and NS--BH merger rates, respectively. Dashed orange lines show the upper and lower limits of the LIGO/Virgo estimated merger rates for all NS--NSs in the local universe. Dashed blue line shows the upper limit of the LIGO/Virgo estimated NS--BH merger rate in the local universe. The LIGO/Virgo estimated merger rates are about 5 orders of magnitude larger than our estimates from GCs. \label{fig:rates}}
\end{center}
\end{figure}

Our local merger rate estimates are summarized in Table~\ref{tab:modelrates} and also shown in Figure~\ref{fig:rates} as a function of redshift. The local rates are calculated by averaging the merger rates for $z < 0.1$ (to eliminate small fluctuations; see inset of lower panel of Figure~\ref{fig:rates}). Assuming the GC density in the local universe is $\rho_{\rm GC}=0.33-2.31\,\rm{Mpc^{-3}}$ \citep{rodriguez2016binary}, we calculate a local merger rate for both NS--NS and NS--BH binaries of $\sim0.009-0.06\,\rm{Gpc^{-3}\,yr^{-1}}$ (Table~\ref{tab:modelrates}), much lower than the local LIGO/Virgo merger rates ($110-3840\,\rm{Gpc^{-3}\,yr^{-1}}$ for NS--NS mergers, and $<610\,\rm{Gpc^{-3}\,yr^{-1}}$ for NS--BH mergers at the 90\% confidence level; \citealp{abbott2019gwtc}). At $z<0.1$, most of these mergers ($\gtrsim90\%$) are from dynamical NS--NS and NS--BH binaries. Only a small fraction of the mergers ($\lesssim10\%$) are from primordial NS--NS and NS--BH binaries (Sec.~\ref{subsec:merger_time}). The rates derived from our extremely optimistic model, coupled with the assumption of a high density, $\rho_{\rm GC}=2.31\,\rm{Mpc^{-3}}$, results in a local merger rate for NS--NS and NS--BH binaries of 25.5~$\rm{Gpc^{-3}\,yr^{-1}}$ and 5.5~$\rm{Gpc^{-3}\,yr^{-1}}$, respectively. These rates are too low in comparison to the LIGO/Virgo rates.

Figure~\ref{fig:rates} shows the merger rates of NS--NS and NS--BH binaries as a function of redshift from our realistic models, assuming a GC density of $\rho_{\rm GC}=0.77\,\rm{Mpc^{-3}}$. We show both the cumulative number of mergers per year, and the volumetric merger rate as a function of redshift. The NS--NS and NS--BH merger rates are comparable at low redshift ($z<1$). Although there are only few BHs remaining in the core at low redshift in the core-collapsed models, they are in general more massive than NSs, and have larger encounter rates. The larger BH encounter rate and the larger number of NSs in the core lead to comparable NS--NS and NS--BH merger rates.

In contrast, there are more NS--NS mergers than NS--BH mergers at high redshift ($z>1$). First of all, there are many more NSs produced through stellar evolution given our assumed IMF, and thus more NS--NS binaries. Most of these binaries are ejected immediately after their formation, and do not have any subsequent dynamical interactions. Furthermore, although there are comparable numbers of NS--NS and NS--BH binaries formed at early times, most of the NS--BH binaries are disrupted (or the NSs are exchanged out of the binaries) following dynamical encounters over short timescales ($<1\,\rm{Gyr}$) before they can inspiral and merge. Of the merged systems, we find that primordial NS--NS and NS--BH binaries, as opposed to dynamical systems, dominate the merger rates at high redshift (Fig.~\ref{fig:alltimes}). This trend does not continue to low redshifts ($z<1$) because most of the primordial systems have already merged (Fig.~\ref{fig:alltimes}).

Directly comparing these results to the current NS--NS and NS--BH merger rate estimates in the local universe from LIGO/Virgo \citep{abbott2019gwtc}, we see that the merger rates from GCs are about 5 orders of magnitude too small. We also note that even our extremely optimistic model, where the number and interaction rate of NSs have been artificially enhanced, still produces merger rates about $1-2$~orders of magnitude below the LIGO/Virgo merger rates.

\newpage
\section{Comparison with Previous Studies}\label{sec:previousstudy}
We now compare our rate estimates to seven previous studies which have examined this question in different ways (summarized in Table~\ref{tab:previousrates}).

\subsection{Theoretical Estimates}\label{subsec:theoretical}
Overall, our NS--NS median merger rate is 10--$10^4$ times lower than previous rate estimates. One natural explanation for this difference is that many previous theoretical studies have used the core-collapsed cluster M15 (the only GC known to contain a NS--NS binary; \citealp{anderson1990discovery,prince1991timing}) as typical for their merger rate calculations. However, M15 is more massive and much denser than typical GCs in the Milky Way \citep{Harris2010catalog}. For instance, \cite{lee2010short} modeled the direct collisions and tidal captures of NSs, as well as binary-single interactions, adopting M15 as a typical background cluster. They concluded that NS--NS mergers in GCs can account for a significant fraction ($>10\%$) of the SGRB rate (Table~\ref{tab:previousrates}).

Later, \cite{bae2014compact} performed direct $N$-body simulations of GCs but used a rather small $N\sim10^4$ for the initial number of stars and employed a simplified IMF with just a few bins. Their merger rate estimate is about 10 times larger than ours, but still very small compared to the LIGO/Virgo merger rate. 

More recently, \cite{belczynski2018origin} derived a NS--NS merger rate from realistic simulations of GCs using the \texttt{MOCCA} code \citep[e.g.,][]{giersz2013mocca}. They used a small number of models with low natal kicks for NSs and only considered NS--NS mergers in local elliptical galaxies (assumed to be about 1/3 of all the galaxies in the local universe). If also taking into account the spiral galaxies, their merger rate estimate can be about $10-20$ times larger than our estimate.

Only one other study has estimated the rate of NS--BH mergers in clusters. \citet{clausen2013black} modeled their formation and merger through binary--single stellar interactions in a static cluster background. They assumed that the clusters retain at most 2 BHs, as in our core-collapsed cluster models.  Nevertheless, their estimate of the NS--BH merger rate is slightly higher than ours (Table~\ref{tab:previousrates}).

\subsection{Empirical Estimates}\label{subsec:empirical}
The earliest empirical NS--NS merger rate estimate \citep{phinney1991rate} took into account three pulsar binaries (including PSR2127+11C in M15), and concluded that GCs have a negligible contribution to the overall NS--NS merger rate in the Milky Way. Later studies (e.g., \citealp{Kalogera2001coalescence}) excluded PSR2127+11C from calculations of the NS--NS empirical merger rate because of its negligible contribution. Overall, the rates presented in this study are 4~orders of magnitude below inferred rates from the observed SGRB luminosity function ($240\,\rm{Gpc^{-3}\,yr^{-1}}$; \citealt{guetta2009short}). Moreover, our realistic (extremely optimistic) rates are about 5 (1--2)~orders of magnitude below the beaming-corrected SGRB event rate ($\sim 100-1000\,\rm{Gpc^{-3}\,yr^{-1}}$; \citealp{fong2015decade,wanderman2015rate}). Most recently, the first two observing runs of LIGO/Virgo have constrained the rates of NS--NS and NS--BH binary mergers in the local universe (Table~\ref{tab:previousrates}; \citealp{abbott2019gwtc}). We find that the LIGO/Virgo NS--NS (NS--BH) rates are about 5 (4)~orders of magnitude larger than our best estimates. These are consistent with the non-detection of a GC to deep limits for the first and closest NS--NS binary merger, GW170817 \citep{fong2019optical}.

Motivated by \cite{andrews2019double}, we also study the orbital periods and eccentricities of all NS--NS binaries ejected from our realistic models. Most of these have large eccentricities ($e>0.5$) and small orbital periods ($P_{orb}<5\,$d). We find that $64\%$ (those with sufficiently large eccentricities and/or small orbital periods) merge within a Hubble time, and the rest ($36\%$, those with relatively large orbital periods) have very long inspiral times. \cite{andrews2019double} suggested that the four observed NS--NS binary pulsars with orbital periods $<1\,$d and eccentricities $>0.5$ may all come from GCs, because their formation is difficult to explain from isolated binary evolution. Our models show that GCs are capable of ejecting NS--NS binary pulsars similar to these four observed systems. However, our estimated NS--NS merger rates from GCs suggest that there is at most 1 NS--NS merger from GCs for every $10^5$ NS--NS mergers in the field. This is clearly inconsistent with the seemingly large fraction (4 out of 20) of observed Galactic NS--NS binary pulsars that \cite{andrews2019double} suggested may have come from GCs.

\begin{deluxetable*}{ccccc}
\tabletypesize{\footnotesize}
\tablewidth{0pt}
\tablecaption{Comparison of Local Volumetric Merger Rates\label{tab:previousrates}}
\tablehead{
\colhead{Type} & \colhead{Rate} & \colhead{Lower Limit} & \colhead{Upper Limit} &\colhead{Reference}\\ 
\colhead{} & \colhead{($\rm{Gpc^{-3}\,yr^{-1}}$)} & \colhead{($\rm{Gpc^{-3}\,yr^{-1}}$)} & \colhead{($\rm{Gpc^{-3}\,yr^{-1}}$)} & \colhead{}
}
\startdata
NS--NS & 0.022 & 0.009 & 0.065 & This study\\
NS--BH & 0.020 & 0.009 & 0.060 & This study\\
\hline
NS--NS & 0.05 & 0.02 & 0.5 & (1)\\
NS--NS & 0.85 & 0.34 & 3.45 & (2)\\
NS--NS & 30 & - & - & (3)\\
NS--NS \& NS--BH & 240 & - & - & (4)\\
NS--NS & 2 & - & - & (5)\\
NS--NS & 1.01 & - & - & (6)\\
NS--BH & 0.03 & 0.01 & 0.17 & (7)\\
\hline
\hline
 & \multicolumn{3}{c}{\it Empirical LIGO/Virgo Rates} & \\
NS--NS & -- & 110 & 3840 & (8)\\
NS--BH & -- & -- & 610 & (8)\\
\enddata
\tablecomments{(1)--\cite{belczynski2018origin}, (2)--\cite{bae2014compact}, (3)--\cite{lee2010short}, (4)--\cite{guetta2009short}, (5)--\cite{grindlay2006short}, (6)--\cite{phinney1991rate}, (7)--\cite{clausen2013black}, (8)--\cite{abbott2019gwtc}. (1)-(7) show merger rate estimates from GCs. Note that the merger rate estimates in \cite{belczynski2018origin} are for GCs in local elliptical galaxies only. Previous studies may use different number densities of GCs in the local universe than ours ($\rho_{\rm GC}=0.77$~Mpc$^{-3}$). Limits from the literature represent the lower and upper bounds given in each work.}
\end{deluxetable*}

\section{DISCUSSION AND SUMMARY}\label{sec:discussions}
Our models for the GC evolution and dynamical interaction of NSs involve a number of theoretical uncertainties. Some of the approximations we make may potentially affect the computed merger rates. We briefly discuss some of these caveats here.

Some previous works have suggested that, under optimistic assumptions about their ability to form long-lived detached binaries (cf. \cite{Kochanek1992tidal,Kumar1996nonlinear}), tidal captures in GCs may increase the formation rates of NS--NS and NS--BH binaries \citep[e.g.,][]{grindlay2006short,lee2010short}. Through tidal captures, additional binaries containing a NS might form in the dense cores of GCs, with companion stars such as main-sequence stars, giants or white dwarfs. Subsequent exchange interactions between these binaries and single NSs (BHs) in the cluster can form NS--NS (NS--BH) binaries that merge within a Hubble time \citep{grindlay2006short,lee2010short}. We do not include tidal captures by NSs or BHs in our models, so our predicted merger rates could in principle be underestimated.

\cite{grindlay2006short} studied the merger rates of NS--NS binaries formed through tidal captures in core-collapsed GCs, and they estimated that roughly 4 merging NS--NS binaries form per Gyr per core-collapsed GC. By assuming that 20$\%$ of GCs are core-collapsed (motivated by observations of Milky Way clusters), they estimated a NS--NS merger rate of 40~$\rm{Gyr^{-1}}$ per 200~GCs in the local universe. Assuming a cluster number density of 0.77~$\rm{Mpc^{-3}}$ in the local universe (as in Sec.~\ref{subsec:rates}), this implies about 0.2~mergers $\rm{Gpc^{-3}\,yr^{-1}}$, which is about 10 times larger than our estimated rate. Thus, in principle, tidal captures might increase significantly the NS--NS merger rate from stellar dynamics. However, we stress that, even under what we regard as optimistic assumptions, the merger rate from GCs is still much lower than the current LIGO/Virgo empirical rate.

For all of the merging NS--NS binaries, $64\%$ of them contain two NSs formed from ECSNe (including accretion-induced collapse and merger-induced collapse since they all receive low natal kicks) and $31\%$ contain one ECSN NS. In total, $95\%$ of the merging NS--NS binaries contain at least one NS formed via ECSNe. On the other hand, about $42\%$ of the merging NS--BH binaries contain an ECSN NS. \cite{Gessner2018hydrodynamical} suggested that NSs formed from ECSNe receive natal kicks of up to a few $\rm{km\,s^{-1}}$ at most, which are lower than the kicks we assumed ($\sigma_{\rm{ECSN}}=20\,\rm{km\,s^{-1}}$; \citealp{Kiel_2008}). For typical clusters with initial $N=8\times10^5$, the escape velocity at early times is about $100\,\rm{km\,s^{-1}}$. Given the large escape velocity, most of the ECSN NSs with $\sigma_{\rm{ECSN}}=20\,\rm{km\,s^{-1}}$ are retained in the clusters. Assuming a few $\rm{km\,s^{-1}}$ for the velocity dispersion instead will slightly increase the NS retention rate, but is not likely to have a large effect on the NS--NS merger rate. Indeed, \cite{belczynski2018origin} used zero natal kicks for ECSN NSs in their NS--NS merger rate calculations and estimated a merger rate of about 10 times higher than ours, but still not high enough to explain the empirical LIGO/Virgo NS--NS merger rate.

Although we have limited this analysis to those globular clusters that survive to the present day, there may also exist a potentially significant number of additional massive clusters that disrupted at earlier times \citep[e.g.,][and references therein]{fragione2018black,rodriguez2018redshift,Krumholz2019star}. Depending on the disruption timescales, these clusters may in principle also contribute to the population of dynamically-formed NS--NS and NS--BH mergers. In total, there are 25 disrupted clusters in our 144 models (Table~\ref{tab:modelgrid}), which are not included in the merger rate calculations. A total of 9 NS--NS and NS--BH mergers (most of them primordial) are found in these disrupted models, where 6 of them are NS--NS mergers and 3 of them are NS--BH mergers. All of them merge in the early universe, and the latest merger occurs after just 4~Gyr. According to \cite{fragione2018black} and \cite{rodriguez2018redshift}, if the total number of NS--NS and NS--BH mergers are affected similarly by including disrupted clusters in the calculations as are BH--BH mergers, this results an enhancement in the NS merger rate by at most a factor of $\sim 2-3$. Given the small number of NS--NS and NS--BH mergers produced by the disrupted clusters, and their early merger times, it seems unlikely that disrupted GCs could have significant contributions to the total merger rates.

In addition, open clusters can also produce NS--NS and NS--BH binary mergers. These clusters are less massive and less dense than GCs, with fewer NSs retained \citep[e.g.,][]{Banerjee2017stellara,Banerjee2017stellarb}, thus the NS--NS and NS--BH merger rates \textit{per cluster} from open clusters are likely even smaller than from GCs. However, previous analyses suggested that the majority of stars may form in low-mass stellar clusters or associations \citep[e.g.,][]{Lada2003embedded}. Taking this into account, young open clusters may in fact contribute significantly overall to the NS--NS and NS--BH merger rates. For example, \citet{Ziosi2014dynamics} showed that young ($t\simeq100\,$Myr) and low-mass ($M_{\rm{cluster}}\approx3500\,M_{\odot}$) clusters can produce a NS--NS merger rate of up to roughly $100\,\rm{Gpc}^{-3}\,\rm{yr}^{-1}$, comparable to the lower end of the empirical LIGO/Virgo rate, assuming $80\%$ of stars are formed in these young star clusters \citep{Lada2003embedded}. 
Although the NS--NS merger rate from young star clusters may be high, it is worth noting that most of the NS--NS binaries from young star clusters are also primordial \citep{Ziosi2014dynamics}, which again stresses that dynamical interactions do not play a significant role in forming NS--NS binaries. Indeed no star cluster was detected at the position of GW170817 down to a limit of $\sim13000\,M_{\odot}$, which rules out $70\%$ of the young massive cluster mass function \citep{fong2019optical}.

We also did not consider dynamical formation of merging binaries in galactic nuclei, which may contribute significantly. Indeed, \cite{petrovich2017greatly} estimated the NS--NS and NS--BH merger rates in galactic nuclei, and found them comparable to our predicted merger rates from GCs.

To summarize, using a large set of realistic models representing Milky Way GCs with a broad range of properties, we have calculated the NS--NS and NS--BH merger rates as a function of redshift. We find that most GCs in the Milky Way retain hundreds of BHs, consistent with other recent studies \citep[e.g.,][]{arca2018mocca,askar2018mocca,Kremer_2018,weatherford2018predicting}. The NSs in these clusters are unlikely to form NS--NS or NS--BH binaries through dynamical interactions. Frequent dynamical interactions of NSs can only happen after most of the BHs have been ejected, which occurs only in the small fraction of core-collapsed GCs (\citealp[e.g.,][]{ye2019millisecond} and Fig.~\ref{fig:alltimes}). 

We have compared our estimates to those of previous studies and to the current LIGO/Virgo empirical merger rates (Sec.~\ref{sec:previousstudy}). We find that the NS--NS and NS--BH merger rates from GCs are negligible compared to the LIGO/Virgo estimates ($\sim10^{-5}$ for NS--NS mergers). We conclude that GCs are not a likely formation {\it or} merger site for merging NS--NS and NS--BH binaries. In order to account for the latest observational constraints on merger rates, other formation channels should be explored, such as binary evolution with a varying common-envelope structure parameter and natal kicks depending on stellar evolution histories \citep{kruckow2018progenitor}, or triple star scenarios \citep{fragione2019blacka,fragione2019blackb,hamers2019double} in the field.

\acknowledgments
We thank Mario Spera, Michael Zevin and
the anonymous referee for useful discussions and comments on the manuscript. This work was supported by NSF Grant AST-1716762 and through the computational resources and staff contributions provided for the {\tt Quest} high performance computing facility at Northwestern University. {\tt Quest} is jointly supported by the Office of the Provost, the Office for Research, and Northwestern University Information Technology. This work also used computing resources at CIERA funded by NSF PHY-1726951.
C.S.Y.\ acknowledges support from NSF Grant DGE-0948017. W.F.\ acknowledges support from NSF Grants AST-1814782 and AST-1909358, NASA Grant HST-GO-15606.001-A from the Space Telescope Science Institute, which is operated by the Association of Universities for Research in Astronomy, Inc., under NASA contract NAS5-26555, and Chandra Award Number G09-20058A issued by the Chandra X-ray Center (operated by the Smithsonian Astrophysical Observatory for and on behalf of NASA under contract NAS8-03060). 
S.C.\ acknowledges support from NASA through Chandra Award Number TM5-16004X issued by the Chandra X-ray Observatory Center
(operated by the Smithsonian Astrophysical Observatory for and on behalf of NASA under contract NAS8-03060). 
G.F. acknowledges support from a CIERA Fellowship at Northwestern University.

\bibliographystyle{aasjournal}
\bibliography{NSbinary_merger_rate}

\appendix
\setcounter{table}{0}
\renewcommand{\thetable}{A\arabic{table}}
\section{Model Properties}
\begin{longrotatetable}
\begin{deluxetable*}{c|ccccccc|ccc|ccc|cc}
\tabletypesize{\scriptsize}
\tablewidth{0pt}
\tablecaption{Cluster Model Properties\label{tab:modelgrid}}
\tablehead{
\colhead{$Model$} & \colhead{$r_c$} & \colhead{$r_{hl}$} & \colhead{$M_{tot}$} & \colhead{$N_{BH}$} & \colhead{$N_{NS}$} & \colhead{$N_{PSR}$} &  \colhead{$N_{MSP}$} & \colhead{$In-cluster$} & \colhead{$Ej$} & \colhead{$EjMerg$} & \colhead{$In-cluster$} & \colhead{$Ej$} & \colhead{$EjMerg$} & \colhead{$N_{NS-NS}$} & \colhead{$N_{NS-BH}$}\\
\colhead{} & \multicolumn{2}{c}{$\rm{pc}$} & \colhead{$10^5\,M_{\odot}$} & \multicolumn{4}{c}{} & \multicolumn{3}{c}{NS--NS at All Times} & \multicolumn{3}{c}{NS--BH at All Times}& \multicolumn{2}{c}{$9<t<12\, Gyr$}
}
\startdata
N2-RV0.5-RG2-Z0.01 & 0.05 & 0.31 & 0.09 & 0 & 24 & 0 & 0 & 0 & 0 & 0 & 1 & 0 & 0 & \multicolumn{2}{c}{disrupted}\\
N4-RV0.5-RG2-Z0.01 & 0.21 & 0.91 & 0.10 & 0 & 104 & 4 & 0 & 0 & 1 & 0 & 0 & 0 & 0 & \multicolumn{2}{c}{disrupted}\\
N8-RV0.5-RG2-Z0.01 & 0.26 & 2.13 & 1.29 & 0 & 655 & 4 & 3 & 2 & 4 & 1 & 0 & 3 & 2 & 10 & 0\\
N16-RV0.5-RG2-Z0.01 & 0.01 & 0.14 & 9.68 & 0 & 0 & 0 & 0 & 0 & 0 & 0 & 0 & 0 & 0 & \multicolumn{2}{c}{disrupted}\\
N2-RV0.5-RG2-Z0.1 & 0.02 & 0.45 & 0.09 & 0 & 12 & 2 & 0 & 0 & 0 & 0 & 0 & 0 & 0 & \multicolumn{2}{c}{disrupted}\\
N4-RV0.5-RG2-Z0.1 & 1.06 & 0.76 & 0.09 & 0 & 36 & 0 & 0 & 0 & 6 & 0 & 0 & 0 & 0 & \multicolumn{2}{c}{disrupted}\\
N8-RV0.5-RG2-Z0.1 & 0.14 & 1.19 & 1.49 & 0 & 259 & 5 & 3 & 1 & 2 & 1 & 1 & 2 & 1 & 3 & 1\\
N16-RV0.5-RG2-Z0.1 & 0.34 & 1.10 & 3.85 & 53 & 733 & 11 & 10 & 0 & 0 & 0 & 0 & 0 & 0 & 0 & 0\\
N2-RV0.5-RG2-Z1.0 & 0.01 & 0.34 & 0.08 & 0 & 11 & 0 & 0 & 0 & 0 & 0 & 0 & 1 & 1 & \multicolumn{2}{c}{disrupted}\\
N4-RV0.5-RG2-Z1.0 & 0.05 & 0.49 & 0.05 & 0 & 42 & 0 & 0 & 0 & 0 & 0 & 0 & 0 & 0 & \multicolumn{2}{c}{disrupted}\\
N8-RV0.5-RG2-Z1.0 & 0.11 & 1.25 & 1.57 & 0 & 285 & 8 & 7 & 0 & 2 & 1 & 0 & 0 & 0 & 0 & 0\\
N16-RV0.5-RG2-Z1.0 & 0.05 & 0.86 & 3.88 & 1 & 749 & 19 & 14 & 1 & 1 & 1 & 2 & 0 & 0 & 2 & 3\\
N2-RV0.5-RG8-Z0.01 & 0.16 & 1.67 & 0.09 & 0 & 39 & 2 & 2 & 0 & 0 & 0 & 0 & 1 & 0 & \multicolumn{2}{c}{disrupted}\\
N4-RV0.5-RG8-Z0.01 & 0.76 & 1.89 & 0.19 & 0 & 94 & 0 & 0 & 0 & 3 & 0 & 0 & 2 & 0 & 2 & 0\\
N8-RV0.5-RG8-Z0.01 & 0.49 & 2.76 & 1.93 & 0 & 778 & 9 & 6 & 3 & 6 & 2 & 2 & 1 & 1 & 18 & 4\\
N16-RV0.5-RG8-Z0.01 & 0.03 & 0.11 & 9.65 & 0 & 0 & 0 & 0 & 0 & 0 & 0 & 0 & 0 & 0 & \multicolumn{2}{c}{disrupted}\\
N2-RV0.5-RG8-Z0.1 & 0.13 & 1.19 & 0.24 & 0 & 16 & 0 & 0 & 0 & 1 & 0 & 0 & 0 & 0 & 0 & 0\\
N4-RV0.5-RG8-Z0.1 & 0.25 & 2.26 & 0.81 & 0 & 87 & 1 & 0 & 0 & 1 & 0 & 1 & 3 & 1 & 1 & 0\\
N8-RV0.5-RG8-Z0.1 & 0.18 & 1.69 & 2.06 & 1 & 283 & 5 & 3 & 1 & 0 & 0 & 1 & 1 & 0 & 5 & 1\\
N16-RV0.5-RG8-Z0.1 & 0.39 & 1.57 & 4.50 & 69 & 752 & 6 & 6 & 0 & 0 & 0 & 0 & 0 & 0 & 0 & 0\\
N2-RV0.5-RG8-Z1.0 & 0.10 & 0.91 & 0.08 & 0 & 6 & 0 & 0 & 0 & 0 & 0 & 0 & 0 & 0 & 0 & 0\\
N4-RV0.5-RG8-Z1.0 & 0.07 & 0.19 & 0.02 & 0 & 30 & 2 & 1 & 0 & 1 & 0 & 0 & 0 & 0 & \multicolumn{2}{c}{disrupted}\\
N8-RV0.5-RG8-Z1.0 & 0.06 & 1.13 & 2.09 & 1 & 265 & 6 & 3 & 0 & 0 & 0 & 2 & 0 & 0 & 5 & 1\\
N16-RV0.5-RG8-Z1.0 & 0.08 & 0.95 & 4.56 & 6 & 751 & 15 & 8 & 1 & 2 & 2 & 2 & 0 & 0 & 3 & 5\\
N2-RV0.5-RG20-Z0.01 & 0.60 & 4.71 & 0.23 & 0 & 39 & 0 & 0 & 0 & 4 & 0 & 0 & 0 & 0 & 1 & 0\\
N4-RV0.5-RG20-Z0.01 & 0.20 & 2.50 & 0.56 & 1 & 171 & 1 & 1 & 0 & 1 & 0 & 3 & 0 & 0 & 0 & 4\\
N8-RV0.5-RG20-Z0.01 & 0.36 & 2.34 & 2.14 & 1 & 838 & 4 & 1 & 2 & 6 & 2 & 1 & 0 & 0 & 15 & 4\\
N16-RV0.5-RG20-Z0.01 & 0.01 & 0.12 & 9.66 & 0 & 0 & 0 & 0 & 0 & 0 & 0 & 0 & 0 & 0 & \multicolumn{2}{c}{disrupted}\\
N2-RV0.5-RG20-Z0.1 & 0.16 & 3.66 & 0.24 & 0 & 12 & 1 & 1 & 0 & 0 & 0 & 0 & 0 & 0 & 0 & 0\\
N4-RV0.5-RG20-Z0.1 & 0.22 & 2.03 & 0.89 & 0 & 82 & 1 & 1 & 0 & 1 & 0 & 0 & 0 & 0 & 2 & 0\\
N8-RV0.5-RG20-Z0.1 & 0.18 & 1.38 & 2.24 & 7 & 311 & 2 & 2 & 0 & 0 & 0 & 0 & 0 & 0 & 0 & 0\\
N16-RV0.5-RG20-Z0.1 & 0.49 & 1.62 & 4.67 & 92 & 812 & 14 & 14 & 0 & 1 & 1 & 0 & 0 & 0 & 0 & 0\\
N2-RV0.5-RG20-Z1.0 & 0.14 & 1.85 & 0.43 & 0 & 20 & 1 & 0 & 0 & 0 & 0 & 0 & 1 & 0 & 1 & 0\\
N4-RV0.5-RG20-Z1.0 & 0.11 & 1.18 & 1.05 & 0 & 86 & 3 & 3 & 0 & 0 & 0 & 0 & 0 & 0 & 0 & 1\\
N8-RV0.5-RG20-Z1.0 & 0.10 & 1.41 & 2.24 & 4 & 281 & 6 & 2 & 0 & 0 & 0 & 1 & 1 & 0 & 3 & 6\\
N16-RV0.5-RG20-Z1.0 & 0.06 & 0.91 & 4.76 & 5 & 749 & 14 & 10 & 0 & 2 & 0 & 2 & 2 & 1 & 2 & 1\\
N2-RV1.0-RG2-Z0.01 & 0.37 & 1.77 & 0.09 & 0 & 26 & 1 & 0 & 0 & 0 & 0 & 0 & 0 & 0 & \multicolumn{2}{c}{disrupted}\\
N4-RV1.0-RG2-Z0.01 & 0.57 & 2.07 & 0.40 & 0 & 143 & 0 & 0 & 0 & 3 & 0 & 0 & 0 & 0 & 1 & 0\\
N8-RV1.0-RG2-Z0.01 & 0.56 & 2.04 & 1.79 & 11 & 677 & 0 & 0 & 0 & 0 & 0 & 0 & 0 & 0 & 0 & 0\\
N16-RV1.0-RG2-Z0.01 & 1.99 & 2.91 & 4.04 & 199 & 1789 & 1 & 1 & 0 & 0 & 0 & 0 & 1 & 0 & 0 & 0\\
N2-RV1.0-RG2-Z0.1 & 0.17 & 1.02 & 0.09 & 0 & 7 & 0 & 0 & 0 & 0 & 0 & 0 & 1 & 0 & \multicolumn{2}{c}{disrupted}\\
N4-RV1.0-RG2-Z0.1 & 0.16 & 1.18 & 0.49 & 0 & 58 & 1 & 1 & 0 & 0 & 0 & 0 & 1 & 1 & 0 & 0\\
N8-RV1.0-RG2-Z0.1 & 0.47 & 1.68 & 1.84 & 28 & 202 & 1 & 1 & 0 & 0 & 0 & 0 & 0 & 0 & 0 & 0\\
N16-RV1.0-RG2-Z0.1 & 1.21 & 2.48 & 4.13 & 197 & 555 & 3 & 3 & 0 & 1 & 1 & 0 & 0 & 0 & 0 & 0\\
N2-RV1.0-RG2-Z1.0 & 0.07 & 0.54 & 0.10 & 0 & 8 & 0 & 0 & 0 & 0 & 0 & 0 & 0 & 0 & \multicolumn{2}{c}{disrupted}\\
N4-RV1.0-RG2-Z1.0 & 0.16 & 0.82 & 0.46 & 1 & 52 & 1 & 1 & 0 & 0 & 0 & 1 & 0 & 0 & 0 & 3\\
N8-RV1.0-RG2-Z1.0 & 0.17 & 0.84 & 1.97 & 10 & 197 & 3 & 2 & 0 & 0 & 0 & 0 & 0 & 0 & 0 & 0\\
N16-RV1.0-RG2-Z1.0 & 0.34 & 1.12 & 4.54 & 131 & 562 & 2 & 2 & 0 & 0 & 0 & 0 & 0 & 0 & 0 & 0\\
N2-RV1.0-RG8-Z0.01 & 0.68 & 2.30 & 0.30 & 0 & 32 & 0 & 0 & 0 & 0 & 0 & 0 & 0 & 0 & 0 & 0\\
N4-RV1.0-RG8-Z0.01 & 0.33 & 2.50 & 0.99 & 1 & 216 & 3 & 2 & 0 & 0 & 0 & 0 & 0 & 0 & 1 & 2\\
N8-RV1.0-RG8-Z0.01 & 0.99 & 2.65 & 2.28 & 34 & 773 & 1 & 1 & 0 & 0 & 0 & 0 & 0 & 0 & 0 & 1\\
N16-RV1.0-RG8-Z0.01 & 1.45 & 3.56 & 4.74 & 240 & 2016 & 4 & 4 & 0 & 0 & 0 & 0 & 0 & 0 & 0 & 0\\
N2-RV1.0-RG8-Z0.1 & 0.08 & 1.80 & 0.39 & 0 & 11 & 0 & 0 & 0 & 0 & 0 & 0 & 0 & 0 & 0 & 0\\
N4-RV1.0-RG8-Z0.1 & 0.36 & 1.69 & 1.07 & 1 & 69 & 0 & 0 & 0 & 0 & 0 & 0 & 0 & 0 & 0 & 0\\
N8-RV1.0-RG8-Z0.1 & 0.69 & 2.04 & 2.26 & 31 & 238 & 0 & 0 & 0 & 0 & 0 & 0 & 0 & 0 & 0 & 0\\
N16-RV1.0-RG8-Z0.1 & 1.27 & 2.66 & 4.70 & 227 & 614 & 4 & 3 & 0 & 0 & 0 & 0 & 0 & 0 & 0 & 0\\
N2-RV1.0-RG8-Z1.0 & 0.06 & 0.94 & 0.45 & 0 & 6 & 0 & 0 & 0 & 0 & 0 & 0 & 1 & 0 & 0 & 0\\
N4-RV1.0-RG8-Z1.0 & 0.09 & 1.38 & 1.08 & 0 & 64 & 2 & 2 & 0 & 0 & 0 & 0 & 0 & 0 & 0 & 0\\
N8-RV1.0-RG8-Z1.0 & 0.22 & 1.05 & 2.37 & 19 & 221 & 2 & 2 & 0 & 0 & 0 & 0 & 0 & 0 & 0 & 1\\
N16-RV1.0-RG8-Z1.0 & 0.42 & 1.24 & 4.97 & 143 & 564 & 2 & 2 & 0 & 0 & 0 & 0 & 0 & 0 & 0 & 1\\
N2-RV1.0-RG20-Z0.01 & 0.39 & 3.19 & 0.45 & 0 & 50 & 0 & 0 & 0 & 0 & 0 & 0 & 1 & 0 & 0 & 0\\
N4-RV1.0-RG20-Z0.01 & 0.72 & 2.45 & 0.94 & 0 & 205 & 2 & 2 & 0 & 1 & 0 & 0 & 2 & 0 & 1 & 0\\
N8-RV1.0-RG20-Z0.01 & 1.01 & 2.89 & 2.40 & 35 & 796 & 0 & 0 & 0 & 0 & 0 & 0 & 1 & 0 & 0 & 1\\
N16-RV1.0-RG20-Z0.01 & 1.69 & 3.98 & 4.94 & 250 & 2051 & 1 & 1 & 0 & 0 & 0 & 0 & 0 & 0 & 0 & 0\\
N2-RV1.0-RG20-Z0.1 & 0.14 & 1.70 & 0.41 & 1 & 13 & 0 & 0 & 0 & 0 & 0 & 0 & 0 & 0 & 0 & 0\\
N4-RV1.0-RG20-Z0.1 & 0.37 & 1.64 & 1.13 & 3 & 82 & 1 & 1 & 0 & 0 & 0 & 0 & 0 & 0 & 0 & 0\\
N8-RV1.0-RG20-Z0.1 & 0.98 & 2.41 & 2.36 & 43 & 245 & 0 & 0 & 0 & 0 & 0 & 0 & 0 & 0 & 0 & 0\\
N16-RV1.0-RG20-Z0.1 & 1.27 & 2.92 & 4.84 & 257 & 618 & 5 & 5 & 0 & 0 & 0 & 0 & 0 & 0 & 0 & 0\\
N2-RV1.0-RG20-Z1.0 & 0.31 & 1.29 & 0.53 & 1 & 19 & 1 & 0 & 0 & 0 & 0 & 0 & 0 & 0 & 0 & 0\\
N4-RV1.0-RG20-Z1.0 & 0.10 & 1.04 & 1.17 & 3 & 77 & 1 & 1 & 0 & 0 & 0 & 0 & 0 & 0 & 0 & 0\\
N8-RV1.0-RG20-Z1.0 & 0.28 & 1.13 & 2.47 & 23 & 219 & 2 & 2 & 0 & 0 & 0 & 0 & 0 & 0 & 0 & 0\\
N16-RV1.0-RG20-Z1.0 & 0.39 & 1.18 & 5.08 & 154 & 577 & 5 & 5 & 0 & 0 & 0 & 0 & 0 & 0 & 0 & 1\\
N2-RV2.0-RG2-Z0.01 & 0.04 & 0.48 & 0.09 & 1 & 9 & 0 & 0 & 0 & 2 & 2 & 0 & 0 & 0 & \multicolumn{2}{c}{disrupted}\\
N4-RV2.0-RG2-Z0.01 & 0.37 & 1.29 & 0.40 & 2 & 90 & 1 & 0 & 0 & 1 & 1 & 0 & 0 & 0 & 0 & 0\\
N8-RV2.0-RG2-Z0.01 & 1.90 & 3.60 & 1.70 & 73 & 425 & 0 & 0 & 0 & 3 & 3 & 0 & 0 & 0 & 0 & 0\\
N16-RV2.0-RG2-Z0.01 & 3.42 & 5.21 & 4.14 & 489 & 1315 & 0 & 0 & 0 & 5 & 5 & 0 & 0 & 0 & 0 & 0\\
N2-RV2.0-RG2-Z0.1 & 0.16 & 0.85 & 0.09 & 0 & 5 & 0 & 0 & 0 & 0 & 0 & 0 & 0 & 0 & \multicolumn{2}{c}{disrupted}\\
N4-RV2.0-RG2-Z0.1 & 0.51 & 1.55 & 0.44 & 1 & 29 & 1 & 1 & 0 & 0 & 0 & 0 & 0 & 0 & 0 & 1\\
N8-RV2.0-RG2-Z0.1 & 3.57 & 3.69 & 1.73 & 95 & 114 & 0 & 0 & 0 & 1 & 1 & 0 & 0 & 0 & 0 & 0\\
N16-RV2.0-RG2-Z0.1 & 2.40 & 4.49 & 4.28 & 497 & 375 & 1 & 1 & 0 & 1 & 1 & 0 & 0 & 0 & 0 & 0\\
N2-RV2.0-RG2-Z1.0 & 0.07 & 0.56 & 0.10 & 0 & 6 & 0 & 0 & 0 & 1 & 1 & 0 & 0 & 0 & \multicolumn{2}{c}{disrupted}\\
N4-RV2.0-RG2-Z1.0 & 0.31 & 1.26 & 0.65 & 16 & 48 & 1 & 1 & 0 & 0 & 0 & 0 & 0 & 0 & 0 & 0\\
N8-RV2.0-RG2-Z1.0 & 0.84 & 2.21 & 2.12 & 116 & 158 & 2 & 2 & 0 & 0 & 0 & 0 & 1 & 0 & 0 & 0\\
N16-RV2.0-RG2-Z1.0 & 1.00 & 2.24 & 4.84 & 433 & 420 & 1 & 1 & 0 & 0 & 0 & 0 & 0 & 0 & 0 & 0\\
N2-RV2.0-RG8-Z0.01 & 0.68 & 2.81 & 0.49 & 0 & 25 & 0 & 0 & 0 & 1 & 1 & 0 & 0 & 0 & 0 & 0\\
N4-RV2.0-RG8-Z0.01 & 1.36 & 3.73 & 1.11 & 13 & 150 & 0 & 0 & 0 & 0 & 0 & 0 & 0 & 0 & 0 & 0\\
N8-RV2.0-RG8-Z0.01 & 1.87 & 4.86 & 2.35 & 112 & 494 & 1 & 1 & 0 & 1 & 1 & 0 & 0 & 0 & 0 & 0\\
N16-RV2.0-RG8-Z0.01 & 3.71 & 5.84 & 4.89 & 610 & 1617 & 1 & 1 & 0 & 0 & 0 & 0 & 0 & 0 & 0 & 0\\
N2-RV2.0-RG8-Z0.1 & 0.25 & 2.56 & 0.48 & 0 & 6 & 0 & 0 & 0 & 0 & 0 & 0 & 0 & 0 & 0 & 0\\
N4-RV2.0-RG8-Z0.1 & 1.43 & 2.85 & 1.10 & 16 & 44 & 0 & 0 & 0 & 0 & 0 & 0 & 0 & 0 & 0 & 0\\
N8-RV2.0-RG8-Z0.1 & 2.00 & 4.28 & 2.35 & 116 & 161 & 1 & 1 & 0 & 0 & 0 & 0 & 0 & 0 & 0 & 0\\
N16-RV2.0-RG8-Z0.1 & 3.44 & 4.54 & 4.84 & 541 & 452 & 3 & 3 & 0 & 1 & 1 & 0 & 0 & 0 & 0 & 0\\
N2-RV2.0-RG8-Z1.0 & 0.57 & 2.27 & 0.56 & 9 & 9 & 0 & 0 & 0 & 0 & 0 & 0 & 0 & 0 & 0 & 0\\
N4-RV2.0-RG8-Z1.0 & 0.90 & 2.52 & 1.21 & 52 & 46 & 0 & 0 & 0 & 0 & 0 & 0 & 0 & 0 & 0 & 0\\
N8-RV2.0-RG8-Z1.0 & 0.83 & 2.28 & 2.52 & 152 & 161 & 2 & 2 & 0 & 0 & 0 & 0 & 0 & 0 & 0 & 0\\
N16-RV2.0-RG8-Z1.0 & 1.04 & 2.39 & 5.15 & 441 & 483 & 2 & 2 & 0 & 1 & 1 & 0 & 0 & 0 & 0 & 0\\
N2-RV2.0-RG20-Z0.01 & 0.75 & 2.83 & 0.54 & 3 & 34 & 0 & 0 & 0 & 2 & 2 & 0 & 1 & 0 & 0 & 0\\
N4-RV2.0-RG20-Z0.01 & 1.70 & 3.21 & 1.20 & 14 & 142 & 0 & 0 & 0 & 0 & 0 & 0 & 0 & 0 & 0 & 0\\
N8-RV2.0-RG20-Z0.01 & 2.56 & 5.00 & 2.46 & 117 & 542 & 0 & 0 & 0 & 4 & 4 & 0 & 0 & 0 & 0 & 0\\
N16-RV2.0-RG20-Z0.01 & 4.92 & 5.74 & 5.08 & 581 & 1673 & 0 & 0 & 0 & 1 & 1 & 0 & 0 & 0 & 0 & 0\\
N2-RV2.0-RG20-Z0.1 & 0.83 & 2.54 & 0.56 & 3 & 6 & 0 & 0 & 0 & 0 & 0 & 0 & 0 & 0 & 0 & 0\\
N4-RV2.0-RG20-Z0.1 & 1.54 & 3.53 & 1.17 & 15 & 47 & 1 & 1 & 0 & 0 & 0 & 0 & 0 & 0 & 0 & 0\\
N8-RV2.0-RG20-Z0.1 & 2.24 & 4.36 & 2.41 & 142 & 171 & 1 & 1 & 0 & 0 & 0 & 0 & 0 & 0 & 0 & 0\\
N16-RV2.0-RG20-Z0.1 & 3.73 & 5.00 & 4.96 & 589 & 451 & 2 & 2 & 0 & 0 & 0 & 0 & 0 & 0 & 0 & 0\\
N2-RV2.0-RG20-Z1.0 & 0.62 & 1.79 & 0.61 & 7 & 18 & 1 & 1 & 0 & 0 & 0 & 0 & 0 & 0 & 0 & 0\\
N4-RV2.0-RG20-Z1.0 & 0.86 & 2.30 & 1.26 & 38 & 52 & 0 & 0 & 0 & 0 & 0 & 0 & 0 & 0 & 0 & 0\\
N8-RV2.0-RG20-Z1.0 & 0.93 & 2.16 & 2.57 & 151 & 167 & 3 & 3 & 0 & 0 & 0 & 0 & 1 & 1 & 0 & 0\\
N16-RV2.0-RG20-Z1.0 & 1.25 & 2.24 & 5.20 & 481 & 490 & 1 & 1 & 0 & 0 & 0 & 0 & 0 & 0 & 0 & 0\\
N2-RV4.0-RG2-Z0.01 & 2.15 & 3.53 & 0.08 & 113 & 5 & 0 & 0 & 0 & 0 & 0 & 0 & 1 & 1 & \multicolumn{2}{c}{disrupted}\\
N4-RV4.0-RG2-Z0.01 & 2.47 & 5.09 & 0.07 & 207 & 28 & 0 & 0 & 0 & 0 & 0 & 0 & 0 & 0 & \multicolumn{2}{c}{disrupted}\\
N8-RV4.0-RG2-Z0.01 & 1369.59 & 6.21 & 0.38 & 438 & 125 & 0 & 0 & 0 & 2 & 2 & 0 & 0 & 0 & \multicolumn{2}{c}{disrupted}\\
N16-RV4.0-RG2-Z0.01 & 8.61 & 8.14 & 3.42 & 892 & 680 & 0 & 0 & 0 & 1 & 1 & 0 & 0 & 0 & 0 & 0\\
N2-RV4.0-RG2-Z0.1 & 3.67 & 3.98 & 0.08 & 110 & 2 & 0 & 0 & 0 & 0 & 0 & 0 & 0 & 0 & \multicolumn{2}{c}{disrupted}\\
N4-RV4.0-RG2-Z0.1 & 3.33 & 5.09 & 0.08 & 202 & 9 & 0 & 0 & 0 & 0 & 0 & 0 & 0 & 0 & \multicolumn{2}{c}{disrupted}\\
N8-RV4.0-RG2-Z0.1 & 7.70 & 6.21 & 0.07 & 351 & 43 & 1 & 1 & 0 & 1 & 1 & 0 & 0 & 0 & \multicolumn{2}{c}{disrupted}\\
N16-RV4.0-RG2-Z0.1 & 7.81 & 7.32 & 3.75 & 907 & 249 & 4 & 4 & 0 & 0 & 0 & 0 & 0 & 0 & 0 & 0\\
N2-RV4.0-RG2-Z1.0 & 1.85 & 2.55 & 0.10 & 76 & 4 & 0 & 0 & 0 & 0 & 0 & 0 & 0 & 0 & \multicolumn{2}{c}{disrupted}\\
N4-RV4.0-RG2-Z1.0 & 1.55 & 2.60 & 0.10 & 72 & 7 & 0 & 0 & 0 & 0 & 0 & 0 & 0 & 0 & \multicolumn{2}{c}{disrupted}\\
N8-RV4.0-RG2-Z1.0 & 2.77 & 3.76 & 1.61 & 300 & 40 & 0 & 0 & 0 & 1 & 1 & 0 & 1 & 1 & 0 & 0\\
N16-RV4.0-RG2-Z1.0 & 2.41 & 4.21 & 4.69 & 816 & 192 & 0 & 0 & 0 & 0 & 0 & 0 & 0 & 0 & 0 & 0\\
N2-RV4.0-RG8-Z0.01 & 1.12 & 3.73 & 0.44 & 6 & 17 & 0 & 0 & 0 & 0 & 0 & 0 & 1 & 1 & 0 & 0\\
N4-RV4.0-RG8-Z0.01 & 4.53 & 7.94 & 1.11 & 53 & 47 & 0 & 0 & 0 & 2 & 2 & 0 & 0 & 0 & 0 & 0\\
N8-RV4.0-RG8-Z0.01 & 7.07 & 7.99 & 2.43 & 330 & 266 & 0 & 0 & 0 & 3 & 3 & 0 & 0 & 0 & 0 & 0\\
N16-RV4.0-RG8-Z0.01 & 6.78 & 8.51 & 5.11 & 1090 & 1073 & 0 & 0 & 0 & 2 & 2 & 1 & 0 & 0 & 0 & 0\\
N2-RV4.0-RG8-Z0.1 & 3.24 & 4.19 & 0.48 & 9 & 4 & 0 & 0 & 0 & 0 & 0 & 0 & 0 & 0 & 0 & 0\\
N4-RV4.0-RG8-Z0.1 & 3.87 & 5.63 & 1.12 & 68 & 21 & 0 & 0 & 0 & 0 & 0 & 0 & 0 & 0 & 0 & 0\\
N8-RV4.0-RG8-Z0.1 & 7.25 & 7.50 & 2.39 & 311 & 75 & 2 & 2 & 0 & 0 & 0 & 0 & 0 & 0 & 0 & 0\\
N16-RV4.0-RG8-Z0.1 & 4.39 & 7.87 & 4.98 & 989 & 338 & 9 & 9 & 0 & 0 & 0 & 0 & 0 & 0 & 0 & 0\\
N2-RV4.0-RG8-Z1.0 & 1.37 & 4.13 & 0.57 & 33 & 2 & 0 & 0 & 0 & 0 & 0 & 0 & 0 & 0 & 0 & 0\\
N4-RV4.0-RG8-Z1.0 & 1.83 & 4.15 & 1.25 & 133 & 18 & 1 & 1 & 0 & 0 & 0 & 0 & 0 & 0 & 0 & 0\\
N8-RV4.0-RG8-Z1.0 & 2.33 & 4.04 & 2.58 & 340 & 70 & 0 & 0 & 0 & 0 & 0 & 0 & 0 & 0 & 0 & 0\\
N16-RV4.0-RG8-Z1.0 & 2.35 & 4.63 & 5.24 & 834 & 258 & 2 & 2 & 0 & 2 & 2 & 0 & 0 & 0 & 0 & 0\\
N2-RV4.0-RG20-Z0.01 & 1.95 & 5.70 & 0.57 & 12 & 16 & 0 & 0 & 0 & 1 & 1 & 0 & 1 & 1 & 0 & 0\\
N4-RV4.0-RG20-Z0.01 & 4.00 & 7.87 & 1.21 & 68 & 75 & 0 & 0 & 0 & 0 & 0 & 0 & 0 & 0 & 0 & 0\\
N8-RV4.0-RG20-Z0.01 & 6.89 & 9.13 & 2.54 & 376 & 272 & 0 & 0 & 0 & 2 & 2 & 0 & 0 & 0 & 0 & 0\\
N16-RV4.0-RG20-Z0.01 & 6.44 & 8.85 & 5.24 & 1107 & 1108 & 1 & 1 & 0 & 1 & 1 & 0 & 0 & 0 & 0 & 0\\
N2-RV4.0-RG20-Z0.1 & 2.19 & 6.42 & 0.57 & 11 & 8 & 1 & 1 & 0 & 0 & 0 & 0 & 0 & 0 & 0 & 0\\
N4-RV4.0-RG20-Z0.1 & 3.35 & 7.15 & 1.18 & 66 & 22 & 2 & 2 & 0 & 1 & 1 & 0 & 0 & 0 & 0 & 0\\
N8-RV4.0-RG20-Z0.1 & 5.82 & 7.73 & 2.48 & 327 & 107 & 5 & 5 & 0 & 0 & 0 & 0 & 0 & 0 & 0 & 0\\
N16-RV4.0-RG20-Z0.1 & 8.54 & 7.99 & 5.11 & 1026 & 346 & 4 & 4 & 0 & 0 & 0 & 0 & 0 & 0 & 0 & 0\\
N2-RV4.0-RG20-Z1.0 & 1.69 & 4.43 & 0.63 & 49 & 5 & 0 & 0 & 0 & 0 & 0 & 0 & 0 & 0 & 0 & 0\\
N4-RV4.0-RG20-Z1.0 & 2.64 & 4.57 & 1.29 & 125 & 19 & 1 & 1 & 0 & 1 & 1 & 0 & 0 & 0 & 0 & 0\\
N8-RV4.0-RG20-Z1.0 & 1.84 & 4.44 & 2.61 & 358 & 85 & 0 & 0 & 0 & 0 & 0 & 0 & 0 & 0 & 0 & 0\\
N16-RV4.0-RG20-Z1.0 & 2.18 & 4.58 & 5.26 & 856 & 259 & 1 & 1 & 0 & 2 & 2 & 0 & 0 & 0 & 0 & 0\\

\enddata
\tablecomments{Column 1: model name (N--initial number of stars in unit of $10^5$; RV--initial virial radius in pc; RG--galactocentric distance in kpc; Z--metallicity in $z_\odot$. For more details on these models, see Kremer et al.\ 2019, Table 4). Columns 2--8: projected core radius, projected half-light radius, total mass, number of BHs, number of NSs, number of all pulsars, and number of MSPs, all at 12~Gyr. Columns 9--11: number of NS--NS mergers in GCs, number of all ejected NS--NS binaries and number of ejected NS--NS binaries that merged within a Hubble time. Columns 12--14: same as columns 9--11 but for NS--BH binaries. Columns 15--16: number of NS--NSs and NS--BHs that appear in GCs between 9 and 12~Gyr.}
\end{deluxetable*}
\end{longrotatetable}

\end{document}